\documentclass[twocolumn,preprintnumbers,amsmath,amssymb]{revtex4}
\usepackage{graphicx}
\usepackage{dcolumn}
\usepackage{textcomp}

\begin{document}

\preprint{}

\title{Threadlike bundle of tubules running inside blood vessels: New anatomical structure}

\author{Xiaowen Jiang}
 \altaffiliation{A permanent address: Dept. of Veterinary, College of Agriculture, Yanbian University, Jilin, China}
\author{Byung-Cheon Lee}
\author{Chunho Choi}
\author{Ku-Youn Baik}
\author{Kwang-Sup Soh}
 \altaffiliation{Correspondence: Kwang-Sup Soh}
 \email{kssoh@kmc.snu.ac.kr}
 \homepage{http://kmc.snu.ac.kr/~kssoh}
\affiliation{Biomedical Physics Lab., School of Physics, Seoul National University, Seoul, 151-747, Korea}

\author{Hee-Kyeong Kim}
\affiliation{School of Chemical Engineering, Seoul National University}

\author{Hak-Soo Shin}
\affiliation{Department of Physics Education, Seoul National University}

\author{Kyung-Soon Soh}
\affiliation{College of Oriental Medicine, Sae Myong University, Chungbook Korea}

\author{Byeung-Soo Cheun}
\affiliation{School of Biochemistry, Inha University, Inchon, Korea}

\date{\today}

\begin{abstract}
According to current anatomy, the arteries and veins do not have threadlike structures running inside the vessels. 
Despite such prevailing knowledge here we report on observation of a novel structure inside the blood vessels of rats and rabbits, which is a semi-transparent elastic bundle of tubules whose diameters are of 10$\mu$m order. 
This is a rediscovery of the Bong Han ducts1,2 which have not been confirmed because the observing method was not known. 
We found a new procedure of observing the intra blood vessel ducts (IBVD) which are too thin, fragile, and semi-transparent to be detected in ordinary surgical operation. 
The method we contrived is to let blood be coagulated around the IBVD so that they become thick and strong by intravenous injection of 10 per cent dextrose solution at the vena femoralis. 
A piece of thickened IBVD sample is treated with urokinase to remove blood clots and the thin thread of IBVD is embedded inside of a string of fibrin.
\end{abstract}

\maketitle

Threadlike structures inside blood vessels of rats and rabbits which are semi-transparent milky white, elastic, and whose diameters are about 50$\mu$m are observed despite current knowledge of animal anatomy and physiology that does not admit such a structure. In this letter we present the method for observing intravenous and intra-arterial threadlike tissues and their tubular substructures.

\begin{figure}
\resizebox{8cm}{!}{
\includegraphics{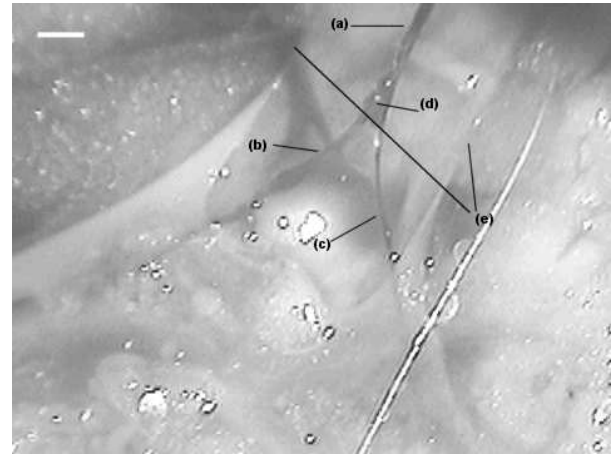}}
\caption{\label{fig1}View of the Intra Blood Vessel Ducts (IBVD) hold tight with a forceps. It looks thick because of coagulated blood clots. It bifurcates at the branching point of the abdominal aorta to the left and right iliac arteries. a, The IBVD. b, c, Two branches of the IBVD. d, The branch point. e, The boundaries of severed vessel walls. Scale bar 0.6mm.}
\end{figure}

A representative sample of the thickened IBVD is shown in Fig.\ref{fig1} where the threadlike structure (a) branches to two smaller ones (b, c) at the bifurcation point (d). It is inside the inferior vena cava of a rat that branches to the left and right iliac vena cava which are incised slowly with a micro dissecting spring scissor (length 105mm, blade 7mm, straight) while the right-side end of the IBVD is held by a micro dissecting forceps (length 100mm,serrated,curved) to keep it tight which is somewhat strong and elastic. 

The existence of the IBVD is not an abnormal state of the particular rat as we have examined nine more cases with similar results without exception, and there are IBVDs in other veins and arteries as we found them at the vessels: the jugular vein, the cardiac vein, the pulmonary vein and artery, the clavicular vein, the ascending aorta, superior vena cava, carotid artery, the hepatic portal vein, the inferior vena cava, the caudal vena cava, the abdominal aorta, the renal veins, the cranial abdominal vein and the right and left iliac veins etc.. The existence of the IBVD is not confined to a rat but probably to all vertebrates including humans. For this we examined four rabbits observing similar IBVDs inside the pulmonary artery, the inferior vena cava, the iliac vein, and even inside the heart (the right atrium). We consider that threads of tubules inside blood vessels are remarkable enough to get wide attention.

\begin{figure}
\resizebox{8cm}{!}
{\includegraphics{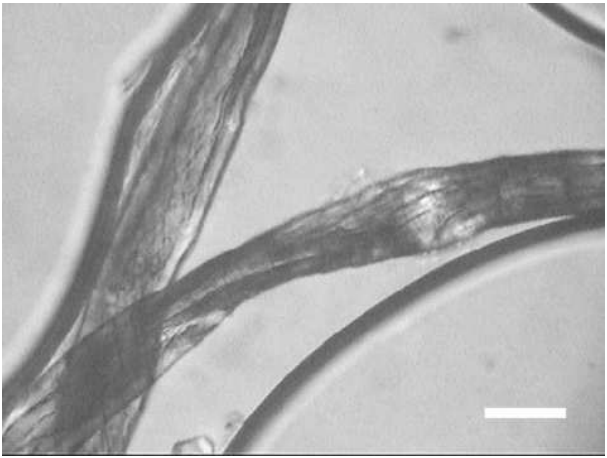}}
\caption{\label{fig2}Two parts of an intra blood vessel ducts cross at the lower left corner inside a water drop with sharp curved boundaries. The IBVD is torn to two pieces at a point ( not seen) which bend to cross each other. The sample is treated with urokinase to remove blood clots. Tubular substructures are seen inside the two ducts. Scale bar 50$\mu$m.}
\end{figure}

Some basic properties of the IBVD are immediately seen. They are soft, elastic, and semi-transparent, their diameters are varying about 50¥ìm, and up to 3cm long we were able to take a sample. The IBVDs looked thick due to the coagulated blood clots around them, which is, in fact, the reason we could see them as in Fig.\ref{fig1}. One vial of fibrinolytic agent urokinase 100,000 I.U. is dissolved in 100ml saline, which is used to dissolve the blood clot at about 37\textcelsius\ for two or three hours. After urokinase treatment the IBVD is observed to be embedded inside a string of fibrins which were precipitated around the 'seed' of the thin IBVD. The IBVD is stained well by methylene blue (0.02\%) 
.The threadlike IBVDs are composed of smaller tubules like a bundle of multiple tubes bound by fibers. Fig.\ref{fig2} shows a good sample of the IBVD after urokinase treated where two threads of an IBVD cross at the lower left corner inside a water-drop with sharp curved boundaries. The IBVD is torn to two pieces which bend to cross each other in this specimen. Tubular substructures are visible inside of the two threads. 

\begin{figure}
\resizebox{8cm}{!}
{\includegraphics{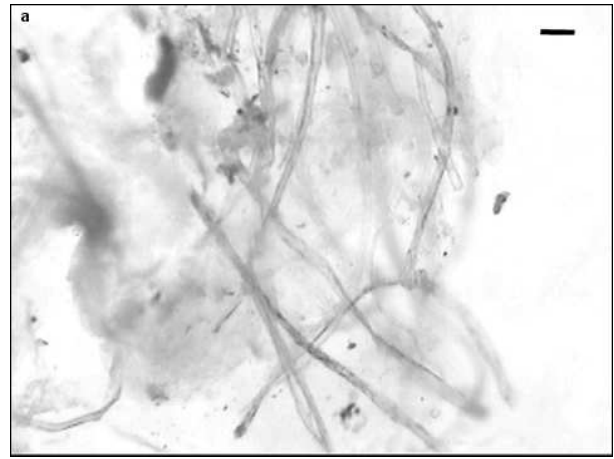}}
\caption{\label{fig3}A sample of the IBVD with blood coagulated is applied with electric field (25V, across 5mm) by two sharp electrodes. The sample is disintegrated slowly, and the scattered tubules appeared after washing with water. Scale bar 50$\mu$m.}
\end{figure}

 It is easy to mistake a string of coagulated fibrins for the IBVD when taking the samples. However, the fibrin-string does not have tubular substructure. Another method of revealing the substructure is to use electrical separation technique. Fig.\ref{fig3} shows scattered tubules on a slide where an electric field (25 V) is applied by two sharp electrodes (two 23 gauge syringe needles, 5mm apart) for few minutes on the sample in a drop of saline water (0.9\%). 
 By repeating the process of applying electric field and water washing one gets the clotted tissue disintegrated to leave the separated tubules. We do not understand the separation mechanism yet.

 It is difficult to find the IBVDs in ordinary surgical operations because they are too thin, semi-transparent and feeble, and shrivel once cut . In the case it is thick due to blood coagulation around it no one would have tried to remove the blood to expose the thin thread. After many failures, we observed the first sample only in the portal vein of a rat , thereafter we examined one rat per day with increasing number of IBVDs observed inside various vessels, and there were some cases even outside of blood vessels. Our search for the IBVD is motivated by the long forgotten theory in early 1960's by Bong Han Kim\cite{Kim} who claimed that not only the IBVD but also a host of similar ducts forms a circulatory network throughout the body whose anatomical and physiological significance are profound and epoch-making as explained in his book in English.\cite{KimBH} His theory has not been confirmed by others despite intense search in 1960's , and the reason for the lack of reproducibility is that he did not disclose the method based upon staining of which no one has yet found the formula and technique. Without such a method it is extremely difficult to search the Bong Han ducts. 

It is our contribution to establish a relatively simple procedure to locate the ducts inside the blood vessels of rats and rabbits, and to observe their substructure of tubules. Novel it may be to find the threadlike duct inside blood vessels, its detail components, physiological functions and the whole network and medical application are far more surprising according to Bong Han theory\cite{Kim,KimBH} which calls for attention from researchers in various fields.

\vspace{0.7cm}
\textbf{Methods}
\vspace{0.4cm}

We obtained rats (Sprague Dawley, 20 weeks old, $\sim$900g) and rabbits(New Zealand White, $\sim$2000g) from the Laboratory Animal Center of Seoul National University with permission Animal Care and Use Commit for searching a circulatory system expected from Traditional Korean Medicine. At 25th minute after a rat is anesthetized an intravenous injection of 10\% dextrose solution is done at the left vena femoralis. At 55th minute about 35ml is injected, and the jugular vein is cut to bleed by itself. At 80th minute the dextrose injection is stopped and the frontal side of the rat is incised. Stomach, intestine, abdominal fat, ribs, and side muscles and flesh are removed such that blood systems are exposed for easy operation. At 130th minute searching of the IBVDs begins from the inferior vena cava and lasts more than two hours taking samples to be examined. The time may vary depending upon the subject animal and the operator's skill.

We do not know the role and mechanism of dextrose solution that happened to be injected for other purpose. But somehow it seems to help the IBVD to be more noticeable, stronger to be pulled out long, and less blood is clotted around.

A variation is to omit the injection of the dextrose water, in which case more blood is clotted around the IBVD and it is less tenacious than in the first method. In the third variation where we omitted both the dextrose injection and the bleeding in jugular vein the IBVD is weakest, easily broken, and more blood is clotted around than the first. Furthermore due to much blood in the vessels it is hard to search the IBVDs. Thus it would be hard to find the IBVD without the help of dextrose injection.

The IBVDs are usually found inside fibrin-strings which are formed with blood coagulated around them, and can be somewhat cleaned up during sampling by heparin saline water(100,000units heparin 25mg, saline 50ml). The searching process was first done with a stereoscopic microscope, but it could be done without it once experienced with few cases. The raw sample is treated with urokinase (100,000units urokinase 1vial at 100ml saline water) to expose the IBVD, or electric field is applied to separate tubules.  

\begin{acknowledgments}
Soh(SNU) is indebted to Professor Marco Bischof for the second reference, and for helps to Drs.Do-Hyun Kim(M.D.), Dae-Hun Park(V.M.D.), Jong-Han Lee(O.M.D.), and Chung-Hyun Lee(V.M.D.).

This work is supported in part by BK21 Korea Research Foundation, Korea Science and Engineering Foundation, Ministry of Industry and Resources, NongHyup SNU branch, and Samsung Advanced Institute of Technology.

Jiang and Kim contributed equally by performing surgery under direction of Soh(SNU) who conducted this work with the research plan made with Shin. Shin found the electrical technique to separate the IBVD ductules. Lee developed the urokinase technique.
\end{acknowledgments}

\bibliographystyle{nature}
\bibliography{tubules}
\end{document}